\documentclass[a4paper,12pt]{article}

\usepackage{amsmath} 
\usepackage{graphicx} 
\usepackage{float} 
\usepackage[round]{natbib} 
\usepackage[hmargin=2.5cm, vmargin=2.5cm]{geometry} 
\usepackage{setspace} 
\usepackage{authblk}

\begin{document}
\onehalfspacing
\title{A double-correlation tremor-location method}
\author[1]{Ka Lok Li}
\author[1,2,3]{Giulia Sgattoni}
\author[1]{Hamzeh Sadeghisorkhani}
\author[1]{Roland Roberts}
\author[1]{\'{O}lafur Gudmundsson}
\affil[1]{Department of Earth Sciences, Uppsala University, Uppsala, Sweden}
\affil[2]{Earth Science Institute, University of Iceland, Reykjavik, Iceland}
\affil[3]{Department of Geology, University of Bologna, Bologna, Italy}

\date{September 23, 2016}
\maketitle
\pagenumbering{arabic}

\begin{abstract}
A double-correlation method is introduced to locate tremor sources based on stacks of complex, doubly-correlated tremor records of multiple triplets of seismographs back projected to hypothetical source locations in a geographic grid. Peaks in the resulting stack of moduli are inferred source locations. The stack of the moduli is a robust measure of energy radiated from a point source or point sources even when the velocity information is imprecise. Application to real data shows how double correlation focuses the source mapping compared to the common single correlation approach. Synthetic tests demonstrate the robustness of the method and its resolution limitations which are controlled by the station geometry, the finite frequency of the signal, the quality of the used velocity information and noise level. Both random noise and signal or noise correlated at time shifts that are inconsistent with the assumed velocity structure can be effectively suppressed. Assuming a surface-wave velocity, we can constrain the source location even if the surface-wave component does not dominate. The method can also in principle be used with body waves in three dimensions, although this requires more data and seismographs placed near the source for depth resolution.
\\ \\
\textbf{keywords}: Interferometry; Persistence, memory, correlations, clustering; Volcano seismology
\end{abstract}

\doublespacing

\section{Introduction}
Many Earth processes, especially in volcanic areas, produce complex seismic signals lacking a clear onset, such as emergent events or continuous tremor. We cannot pick their onsets in order to locate their sources. Instead, amplitude decay with distance has sometimes been used \citep[e.g.][]{Battaglia2003,DiGrazia2006}. This requires simplistic assumptions about geometrical spreading, site magnification and attenuation and does not work well for the tremor data at Katla volcano, which we use as an example in this paper (Sgattoni et al., submitted manuscript). Another strategy is to use array methods, such as beam forming \citep{Rost2002}, FK analysis \citep{Capon1973} or semblance methods \citep{Furumoto1992,Almendros2003,Konstantinou2003} on station sub-arrays. This requires dense sampling of the wave field to yield reliable direction and slowness or vectorial wave number. Combining such estimates at multiple dense arrays constrains the source location. In classical beam forming \citep{Rost2002}, after appropriate delay that corresponds to a given trial source location, data series from all stations are summed, the result squared, and this summed over time. The result is identical to the sum of all autocorrelations and appropriately delayed cross correlations of the individual time series. The autocorrelations add noise, but contain no timing information, so excluding these and summing only the cross correlations helps suppress noise without loss of relative time information.

Inter-station cross-correlation has become a common tool for analysis of seismic ambient noise over the past decade. Numerous authors have used cross-correlation analysis in order to assess the spatial distribution of ambient noise sources. \citet{Shapiro2006} located a peak near the primary microseismic peak at 26 s period in the Gulf of Guinea by back projecting noise correlations. \citet{Gudmundsson2010}, \citet{Zeng2010}, \citet{Ballmer2013} and \citet{Droznin2015} used a similar approach to locate persistent sources of geothermal noise in Iceland, microseisms around Japan and volcanic tremor in Hawaii and Kamchatka, respectively. \citet{Haney2010} measured differential travel-times extracted from cross-correlations of recordings of very-long-period (VLP) volcanic tremor at different stations and inverted those for the source location. All such methods require detailed information about seismic velocity in the area. Because this information is often insufficiently precise, envelopes or various integral measures of cross correlations were stacked rather than the cross correlations themselves in a number of the above studies.

If we think of a tremor or noise source as a point source which is incoherent in time and consider a simple medium between that source and all recording seismographs, then the recorded tremor (noise) will consist of a continuous interference pattern between the same source functions at random time intervals, but at a fixed delay for one station compared to any other corresponding to the differential travel-time from the source to the two stations. Recordings of the tremor at two stations will, therefore, correlate at that time lag, cancelling the random time of each source pulse by a relative measurement. This is the principle that all of the above methods make use of. In addition, correlation is an effective tool to suppress noise, or any signal that is incoherent at the particular time lags assumed.

Each differential travel-time extracted by correlation of tremor recorded at two stations does not constrain the source location to a point alone. Instead, it will constrain the location to lie anywhere on a hyperbolic surface corresponding to equal differential distance from the source (assuming a uniform velocity). If we restrict the propagation to two dimensions the solutions are constrained to lie on a hyperbolic curve. With information from two station pairs, we have two hyperbolic curves, which will usually intersect at one or two points. Further station pairs will provide redundant constraints on a single-point solution. Thus, we need differential travel-time measurements from at least two station pairs to locate a source in
two dimensions. Similarly, we need three pairs to locate a source in
three dimensions.

If we back project correlation functions to hypothetical source locations, again using uniform velocity, the energy distribution for each of them will peak on a hyperbolic surface. Stacking two correlation functions will usually produce a peak at a point, while stacking more station pairs will enhance that peak relative to other parts of the hyperbolae. However, this suppression is rather weak and is fundamentally limited by the number of available station pairs. The method we present below is intended to enhance signal by using a modified form of correlation, which we refer to as double correlation. The double correlation calculates the cross correlation of correlations from two pairs of stations. It uses the fact that when the signals are temporarily aligned for a given source, they should correlate not only between stations, but also between cross-correlation functions. The double correlation will suppress noise correlated at time shifts that are inconsistent with the assumed velocity model. It will also suppress the random noise in the signals.

Various assumptions are implicit in all correlation approaches, including that a source is not significantly autocorrelated and that it radiates similar time series in all directions. Besides, we need to know the velocity well, which we may not. The tremor signals may contain both body and surface waves, so it becomes ambiguous which velocity to use and if we need to solve a two-dimensional or a three-dimensional problem. The Green's functions between the source and different receivers may be complex and differ by more than a time delay from one station to the other (multipathing, phase conversion). In that case the covariograms may correlate poorly. However, variations in the tremor source activity level should still produce similar variations at each station pair in the more slowly-varying covariogram envelopes. Use of envelopes may help to stabilize the back projection at a loss of spatial resolution.

\section{Method}
Our method of location extends the previous correlation methods by applying double correlation instead of a single correlation. We make the same assumptions as are normally made in, e.g., single-correlation methods, including that if multiple sources are present, they are not correlated with each other. The double-correlation location method can be divided into three aspects: (1) the computation of the double correlations of triplets of seismographs, (2) the back projection of the stacked double correlation to hypothetical sources in a geographic grid, and (3) the selection of station triplets. In the following subsections we address these aspects separately.

\subsection{Double correlation}
Here we introduce the procedure to compute a double correlation of a station triplet. The correlation at a pair of stations, $a$ and $b$, is
\begin{eqnarray}
C_{ab}(j) = \sum\limits_{i=1}^{N-j} a(i) b(i+j) ,
\end{eqnarray}
where $i$ and $j$ are sample indices, $a(i)$ and $b(i)$ are the seismograms at station $a$ and $b$, respectively, and $N$ is the total number of sample points. Having computed a second such correlation for stations $c$ and $d$
\begin{eqnarray}
C_{cd}(j) = \sum\limits_{i=1}^{N-j} c(i) d(i+j) ,
\end{eqnarray}
we can then define a double correlation as
\begin{eqnarray}
\label{eq:double_xcorr}
C_{abcd}(m) = \sum\limits_{j=-M}^{M} C_{ab}(j) C_{cd}(j+m) ,
\end{eqnarray}
where $M = N - 1$ is the maximum sample lag in the first cross correlation. The double correlation is useful for two purposes. First, the level of redundancy is increased over a single correlation, which suppresses random noise. Second, this double correlation will peak at the double differential travel-time between the two pairs and if we back project this function to the source in two dimensions it will concentrate energy around a point as opposed to the hyperbola obtained by the back projection of a single correlation.

This is, however, not how we implement double correlation. We note that the sum in eq. (\ref{eq:double_xcorr}) will not have much of a contribution from time lags beyond a finite limit corresponding to the possible differential travel-times of sources within the network. Therefore, we implement double correlation in the following manner. First, we convert our time series to an analytic signal
\begin{eqnarray}
\hat{f}(t) = f(t) + i \mathcal{H}\big[f(t)\big] ,
\end{eqnarray}
where $f(t)$ is the original time series, $\mathcal{H}\big[...\big]$ denotes the Hilbert transform and $\hat{f}(t)$ is the analytic signal of $f(t)$. Then, we divide the time span of our signals, $T$, into $K$ equal sub intervals of length $\Delta T = T / K$. Next, we select a triplet of stations, $a$, $b$ and $c$. We choose one of those as a reference station, for example $a$, and correlate stations $a$ and $b$ on one hand and stations $a$ and $c$ on the other. Choosing this particular way of combining station is not fundamentally required for our approach to work, but for our data character and geometry, this approach is effective. The issue related to the selection of station triplets will be discussed later. For the $k$-th interval the correlations at station $a$ and $b$, and station $a$ and $c$ are
\begin{eqnarray}
{}_k\hat{C}_{ab}(j) = \sum\limits_{i=k\Delta T+1}^{(k+1)\Delta T} \hat{a}(i) \hat{b}^*(i+j)
\end{eqnarray}
and
\begin{eqnarray}
{}_k\hat{C}_{ac}(j) = \sum\limits_{i=k\Delta T+1}^{(k+1)\Delta T} \hat{a}(i) \hat{c}^*(i+j),
\end{eqnarray}
respectively, where the time series are all complex and $^*$ indicates the complex conjugate. Finally, we correlate those two
\begin{eqnarray}
\hat{C}_{abc}(m) = \sum\limits_{k=0}^{K-1} {}_k\hat{C}_{ab}(i) \thinspace \thinspace {}_k\hat{C}_{ac}^*(j+m) ,
\end{eqnarray}
where $i$ and $j$ are the expected sample lags calculated from the theoretical differential travel-times between station $a$ and $b$, and $a$ and $c$, respectively. These theoretical differential travel-times are obtained from, e.g., ray tracing in a pre-defined velocity model.

The amplitude of this complex double correlation $\hat{C}_{abc}$ is a measure of the double correlation at a given differential time-shift. It will peak at the double differential travel-time between the two station pairs. The phase of $\hat{C}_{abc}$ measures the precise phase lag or time lag between the time series. In order to make use of the phase information in $\hat{C}_{abc}$, a complete velocity model, which we may not know precisely, is needed to calculate the differential travel-times between stations. We, therefore, focus our analysis on the amplitude of $\hat{C}_{abc}$.

The double correlation will suppress random noise more than single correlation and suppress non-stationary correlated noise. Stacking the back-projected double correlations for all triplets will further suppress correlated peaks that are not common for all triplets, e.g., components of the Green's functions between source and stations that vary from one path to another. A further advantage of this double correlation is that each back projection provides a localized estimate of the source location (in two dimensions). Therefore, when we back project to a hypothetical source and stack different such double correlations, we stack equivalent estimates of the source location. However, stacking single correlations involves stacking hyperbolic distributions that have very different localization properties.

\subsection{Back projection}
The back projection of the above double correlation is done in the following manner: For each hypothetical source location the travel time to any given station is computed using a velocity model. The time lags between stations $a$ and $b$ and stations $a$ and $c$ are thus estimated and their double difference computed. The value of the back projection is then the value of the corresponding double correlation at that differential lag. These are then stacked over all triplets.

It is common that amplitude variations of seismograms are not compensated in the back projection. The amplitude variation depends on a number of effects, e.g., focusing effects due to velocity variations, attenuation etc. We argue that for Katla volcano, we cannot know the amplitude variation sufficiently well since we are not able to model these effects precisely. Besides, the amplitude variation is not important, because we are exploiting coherent phase information in our method. The amplitude simply affects the relative weighting of the different covariograms or double covariograms. In order to optimize the use of the phase information, it is more appropriate to select such weighting based on some measure of correlation uncertainties in each case. However, such uncertainties may be difficult to estimate reliably and will vary from one data set to the other.

\subsection{Selection of station triplets}
In the calculation of double correlations we choose to combine stations according to a reference-station scheme, rather than including all possible combinations of stations. We have tested this synthetically, also using the same real-data examples as presented later. We find that this reference-station scheme works better than including all possibilities. One reason is that by restricting the double correlations to correlations within closed triangles reduces the occurrence of nearly parallel station pairs. It, in turn, alleviates overlapping of tangential hyperbolae. As a result, hyperbolic streaking is reduced in the result.

Under this reference-station scheme, a network of $n$ stations will have $C^n_3$ combinations of three stations and for each, three possible choices of a reference station. The order of the remaining two stations in each triplet is arbitrary and the two choices are completely redundant. There are, therefore, $N$ independent triplets, where $N$ is
\begin{eqnarray}
N = 3 C^n_3 .
\end{eqnarray}
For a station network with $n = 10$ seismographs the number of triplets is $N = 360$.

\section{Application to real data}
\begin{figure}[H]
\centering
\includegraphics{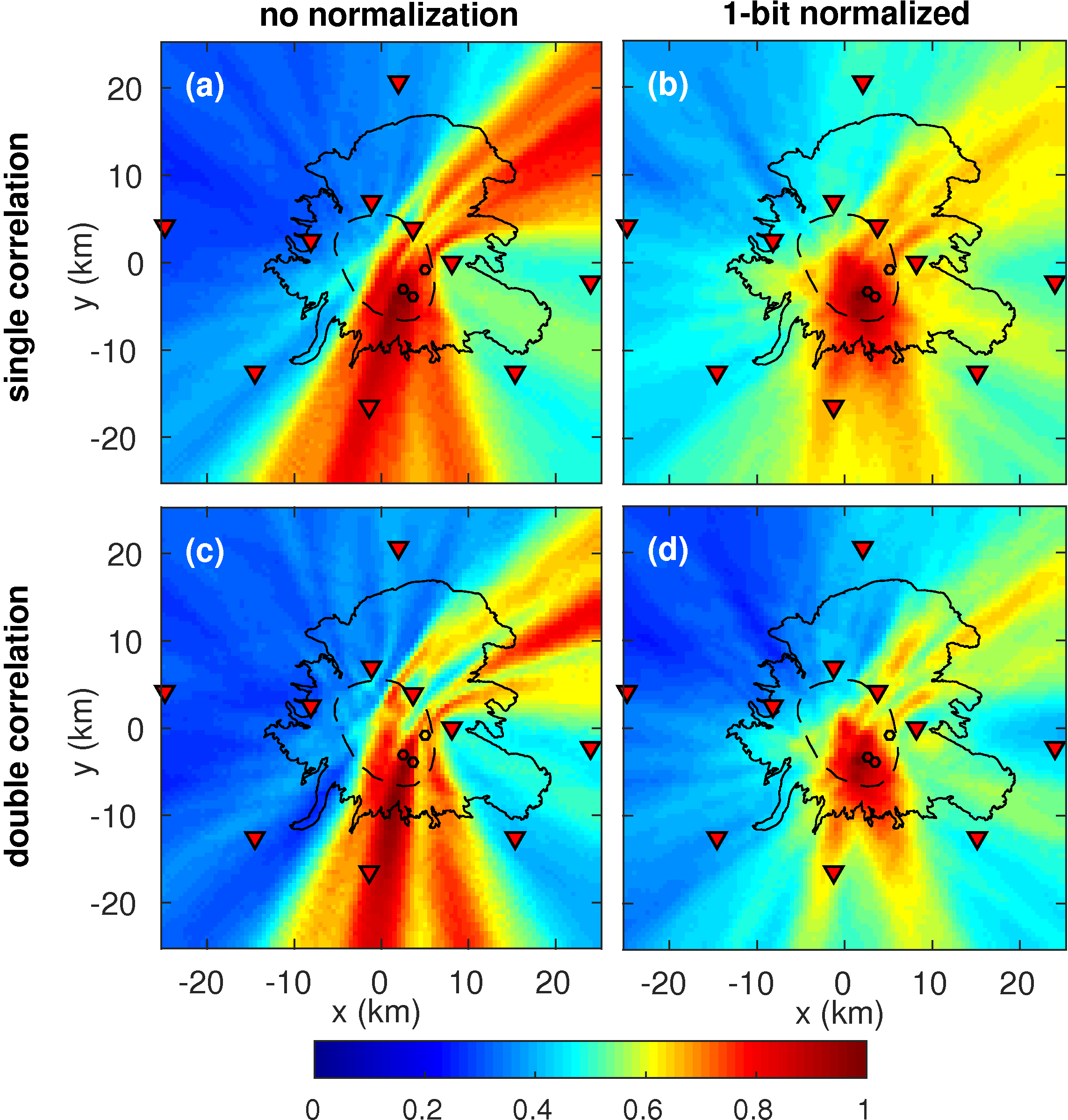}
\caption{(a) and (b): The back projection of single correlations for a tremor at Katla volcano, Iceland in July 2011. Two strategies of amplitude normalization are used. From left to right: without any normalization and one-bit normalization of seismograms. (c) and (d): the same as (a) and (b) except showing the back projection of double correlations. The square root of back-projected double correlations is taken to allow a fair comparison to the single correlation. The colours define normalised energy, dark red for maximum energy. Red inverted triangles: seismic stations. Black solid line: glacier. Black dashed line: caldera outline. Black open circles: locations of cauldrons.}
\label{fig:katla}
\end{figure}

We apply the double-correlation method to tremor at Katla volcano, southern Iceland, during unrest in July 2011. Katla volcano is located near the central south coast of Iceland near the tip of the southward propagating Eastern Volcanic Zone (Fig. 1 in Supporting Information). This is a sub-glacial volcano with a well developed caldera structure (about 10 km in diameter). The geology of the area is complex, indicating that there could be strong attenuation and focusing effects. Both the ice cover and the caldera rim are outlined in Figs \ref{fig:katla} and \ref{fig:synthetic} for reference. The unrest of the volcano started in August 2010 and an increase of seismicity in the area was observed in July 2011. The tremor started at $\sim$ 19:00 GMT on July 8th 2011 and lasted for 23 hours (see Sgattoni et al., submitted manuscript for a detailed description of this unrest). Three main cauldrons (depression on the glacier surface) collapsed on the ice surface during the tremor \citep{Gudmundsson2013}. Those are also indicated in Fig. \ref{fig:katla} for reference. Including the permanent stations from the Icelandic Meteorological Office and three temporary stations deployed by Uppsala University and Reykjavik University on nunataks (rock protruding the ice surface) in the ice, a total of 12 nearby stations were in operation during the tremor period. Here we use the 10 stations closest to the caldera for our analysis.

First, local earthquakes are removed from the seismograms by automated clipping and manual removal. This is not because events invalidate our approach, but because they may not be at the same location as the tremor source. As the events are strong signals, they may thus inject strong ``correlated noise'' to the analysis of tremor. Data are then filtered between 0.8 and 1.5 Hz using a 4th order Butterworth filter. We choose this low-frequency band because the energy is more stable there than at higher frequencies (Sgattoni et al., submitted manuscript). We expect that the tremor sources are located inside the station network and the amplitude of some of the closer stations may dominate. It is, therefore, sensible to normalize amplitudes in some way. We compare two normalization strategies: (1) using raw covariograms without any normalization of traces; (2) using covariograms after one-bit normalization of each trace. These choices are subjective. To optimally weigh the covariograms in the back-projected stack, we need to know their uncertainties, which are difficult to estimate.

Here we focus on the surface-wave component of signals and, therefore, perform the back projection in a two-dimensional grid with a resolution of 0.5 km in each direction. We have tested several uniform velocities and find that a velocity of 1.2 km/s best focuses the energy. Since we are working with the modulus of the complex double correlation, which is related to the surface-wave envelopes (energy), this velocity should correspond to group velocity. This may appear to be a low velocity. However, it is comparable to the results of \citet{Haney2011} at Hekla volcano and \citet{Obermann2016} at Sn\ae fellsj\"okull volcano, as well as Jeddi et al. (manuscript in preparation) at Katla volcano and Benediktsd\'ottir et al. (manuscript in preparation) at Eyjafjallaj\"okull, all in Iceland, at similar frequencies (1 Hz).

Fig. \ref{fig:katla} shows the back projections of single and double correlations using the different normalization strategies. For case 1 (without any normalization) the energy peaks along hyperbolae corresponding to constant time shifts for the three aligned stations along the northeastern caldera rim. This energy is significantly better focused in the double correlation than the single correlation results. However, an elongate streak of energy persists to the south of the caldera. The background energy is also lowered in relative terms by double correlation compared to the single correlation, demonstrating the effect of noise suppression. In case 2 (correlating one-bit normalized traces) the hyperbolic streaking artefact is significantly suppressed. Again, the double correlation suppresses these artefacts more effectively than the single correlation.

Note, that we are not able to evaluate which normalization scheme is optimal unless we know the characteristics of noise processes in the signals. In our particular examples (real and synthetic data, see also Fig. \ref{fig:synthetic}) case 2 appears to be the better choice. Case 1 may be suboptimal because the high-amplitude stations near the source dominate and we, therefore, do not take full advantage of all possible geometries offered by the network.

In all cases, the maximum back-projected energy lies close to the location of two closely spaced cauldrons that collapsed in the southern part of the caldera. This lends credibility to the location method, because it is reasonable to expect tremor in association with volcanic, geothermal or glacio-hydrological processes near cauldrons. The third cauldron appears to be associated with weaker tremor (Sgattoni et al., submitted manuscript).

\section{Synthetic tests}
We have tested the double correlation approach with various synthetic tests. When using perfect data in a uniform velocity structure, the single source is well recovered with a resolution limitation controlled by the frequency content of the data. With significant velocity heterogeneity (compared to the finite frequency limitation) we lose our ability to predict travel time precisely. This results in erratic phase behaviour in the complex covariograms. However, the modulus of the back-projected double correlation remains stable, but its finite width, controlled by the frequency band-width of the data, translates into a broader location estimate.

Our tests of the method include both various noise processes and propagation effects. We have included random incoherent noise in the time series. We have added coherent noise in the form of the source impulses delayed according to a simple body-wave model to simulate the presence of body waves in the tremor signal. We have added signals corresponding to plane waves from a range of azimuths to simulate the effect of distant sources, e.g. microseisms. Finally, we have added a set of random scatterers given a random strength, a random scattering width and a random orientation. We have also modelled the effect of heterogeneity by delaying the source impulses according to a two-dimensional velocity model for surface waves and given those random amplitude perturbations on top of simple geometrical spreading. The tests are done with seismograms filtered in the same frequency range as the Katla tremor data (0.8-1.5 Hz). The back projection is done with the same uniform velocity as chosen for the Katla tremor data, which is also the average group velocity in the synthetic velocity structure.

We have done a number of tests with different strength of the described effects added. These tests reveal that, even with a strong velocity variation of $\pm$ 20 \% (root-mean-square level of heterogeneity) with a correlation length of 7 km, we still recover the source location to within a few km. Suppression of incoherent noise is very effective and the point-source location is well recovered even if this noise has the same amplitude as the signal on average (signal-to-noise ratio is 1). The body-wave component of the synthetic seismograms is coherent at time delays corresponding to distance and velocities that are very different from those for the surface waves and does not significantly affect the recovery of the point source even if given a relative amplitude as big as the surface-wave component. Despite the appearance of some hyperbolic artefacts, distant plane-wave sources have little effect on the source location estimate unless they dominate over the local source in amplitude. The most significant synthetic components are the addition of scatterers and random amplitude variations. With 70 scatterers randomly distributed within the caldera, given a random orientation and a Gaussian width of $\pm$ 60$^{\circ}$ (one standard deviation) and a random scattering strength between 0 and 1, we simulate the appearance of the stacked, back-projected energy distribution obtained with the real data. By enhancing the amplitudes of synthetic recordings at stations on nunataks in the ice cap we simulate the effects of different normalization strategies on the back-projected distribution. Note that the geology in Katla volcano is very complex so that the site magnification and attenuation effects could be strong. This would result in dramatic variations in amplitudes between different seismograms.

\begin{figure}[H]
\centering
\includegraphics{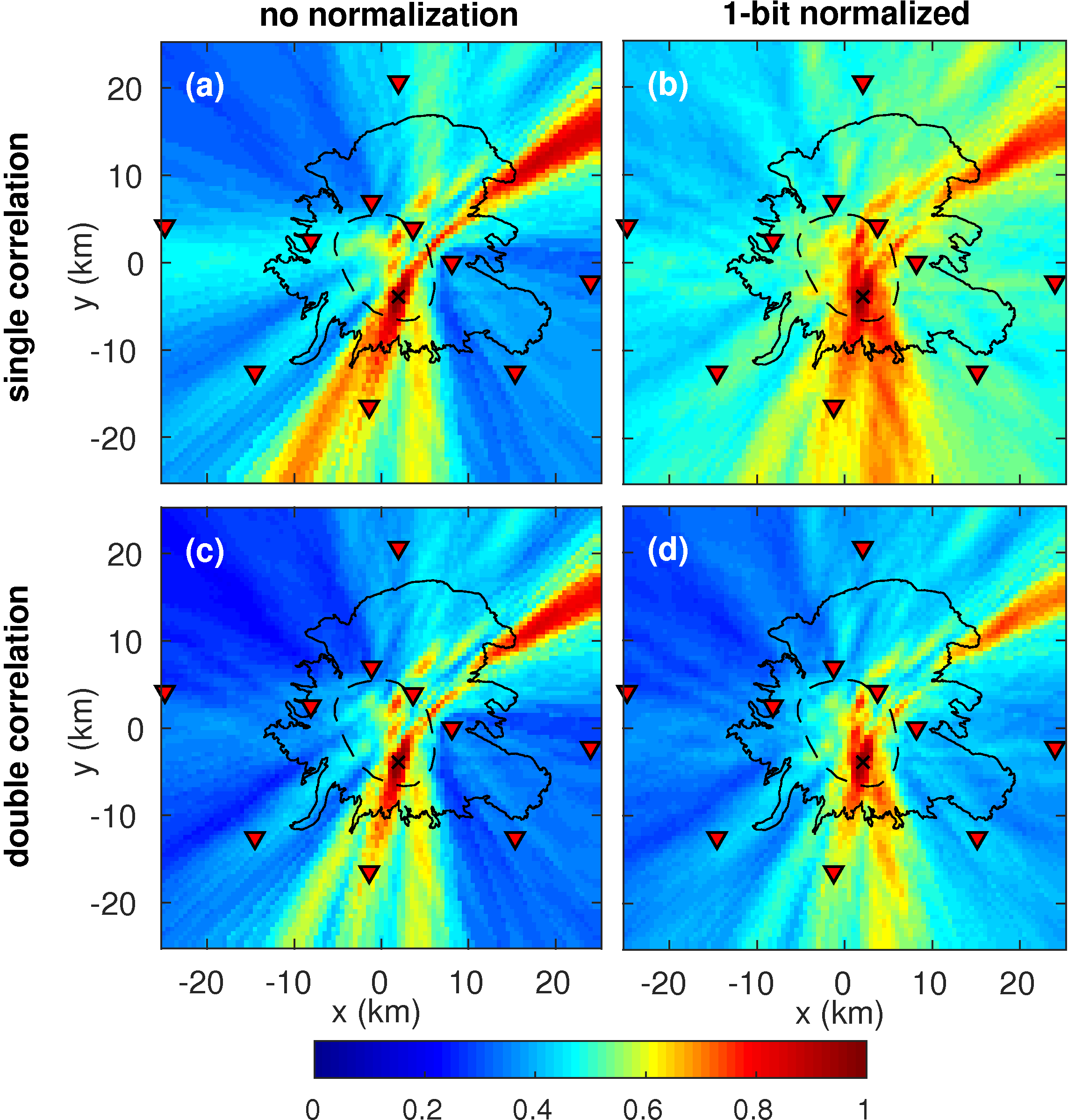}
\caption{(a) and (b): The back projection of single correlations for synthetic examples. Two strategies of amplitude normalization are used. From left to right: without any normalization and one-bit normalization of seismograms. (c) and (d): the same as (a) and (b) except showing the back projection of double correlations, again the square root is taken for the double correlations. For better comparison the same colour scale is used as the real-data example. The symbols follow Fig. \ref{fig:katla}, except that the black cross denotes the source location.}
\label{fig:synthetic}
\end{figure}

In the testing we also try to simulate the features appeared in the real-data example. Four examples of such simulations are shown in Fig. \ref{fig:synthetic}. They include both single and double correlation, again with two different ways of amplitude normalizations. In all cases we use synthetic data containing velocity heterogeneity ($\pm$ 20 \% velocity variation, 7 km correlation length), added white noise (signal-to-noise ratio of 1 on average), a body-wave component equal to the surface-wave component in amplitude and a plane-wave component with a relative amplitude of 0.1 on average. We have added 70 random scatterers with the same geometrical characteristics as described above. In Figs \ref{fig:synthetic}(a) and (c) we show the back-projected energy without any amplitude normalization. The source, which is located at coordinates (2, -4), is clearly resolved (marked with a black cross in Fig. \ref{fig:synthetic}). These synthetic tests show similar behaviour as the real-data examples in which the strong spurious peaks are suppressed in the double correlation. The energy is also more focused by double correlation than single correlation. Note the common structure in the peripheral distribution, which is controlled by the station geometry. Figs \ref{fig:synthetic}(b) and (d) show the back-projected maps with one-bit normalization of traces. Compared to Figs \ref{fig:synthetic}(a) and (c) the streaking hyperbolic artefacts are reduced. The energy is more focused around the central source region in the double- than the single correlation.

\section{Discussion and conclusions}
We introduce a double-correlation method to locate tremor sources. This method back projects stacks of complex, doubly correlated tremor records to hypothetical sources in a geographic grid for location estimates. Compared to the existing single-correlation method, the double correlation is better in suppressing both random noise and noise correlated at time shifts that are inconsistent with the assumed velocity. The back projection of each double correlation provides a localized estimate of source locations. Stacking different back-projected maps will, therefore, give equivalent estimates of the source location, as opposed to the single correlation, where the individual maps contain strong signatures of the station geometry.

Application of the method to real data from Katla volcano, Iceland reveals that there is a stable energy source during the tremor period. The location of that source lies close to the cauldrons collapsed during that period.

In the synthetic tests we are able to recover the primary source location with a resolution of a few km despite various types of noise added to the seismograms. Although we have added significant secondary sources, such as scatterers and plane-wave sources, these are well suppressed by the double correlation. Unless they are comparable in amplitude to the primary source, they will not have a significant effect on its source-location estimate. Random noise is effectively suppressed by the double correlation. Correlated noise, such as the body-wave component, does not have a significant effect on the result because it has a time delay which is very different from the surface-wave component. Velocity perturbation has a little effect on the location estimate except broadening it to have a resolution of a few km, depending on the frequency content of the data.

We are able to simulate the real-data example qualitatively with the added effects to the synthetic data. The energy distributions in the back-projected maps are similar for both data sets. In particular, hyperbolic streaking artefacts in the real data are well simulated. We, therefore, conclude that the peak in Fig. \ref{fig:katla}(d) reveals the location of the dominating energy source during the tremor period.

We need not assume that the tremor is dominantly composed of surface waves for a two-dimensional back-projection to work as at surface-wave velocities the body-wave components do not stack coherently. It suffices that the tremor contains a significant surface-wave component. Focusing on surface waves limits the analysis to two dimensions, so source depth cannot be recovered. If the source has a finite depth, a phase shift will be introduced into the surface waves, but this will be common to all observations and, therefore, cancel in the double-differential times of the double correlation.

For the higher frequencies (1.5 - 4 Hz and 4 - 9 Hz) of the Katla data we have not been able to recover a coherent picture of the source with this method. We suspect that is caused by increasing vigour of scattering effects at increasing frequency in tune with the fact that addition of discrete scatterers for the 0.8 - 1.5 Hz range proved to be a significant obscuring effect in our synthesized examples.

In this short article, we have not discussed all possible complications, such as that for some station geometries with multiple dominating sources some spurious correlation may be incompletely suppressed even after double correlation. Such problems may be alleviated by further processing steps, e.g. with higher-order cross correlations. The specific method presented here is only one of a family of methods based on higher-level cross correlation. Slightly different approaches may be better suited depending on the characteristics of the data. However, the suggested double-correlation strategy does better suppress some types of uncorrelated and correlated noise than other alternatives and may be useful for a spectrum of array-processing problems, not only tremor analysis.

\bibliographystyle{plainnat}
\bibliography{double_corr}

\section*{Supporting information}
\renewcommand{\thefigure}{S\arabic{figure}}
\setcounter{figure}{0}
\begin{figure}[H]
\centering
\includegraphics[width=26pc]{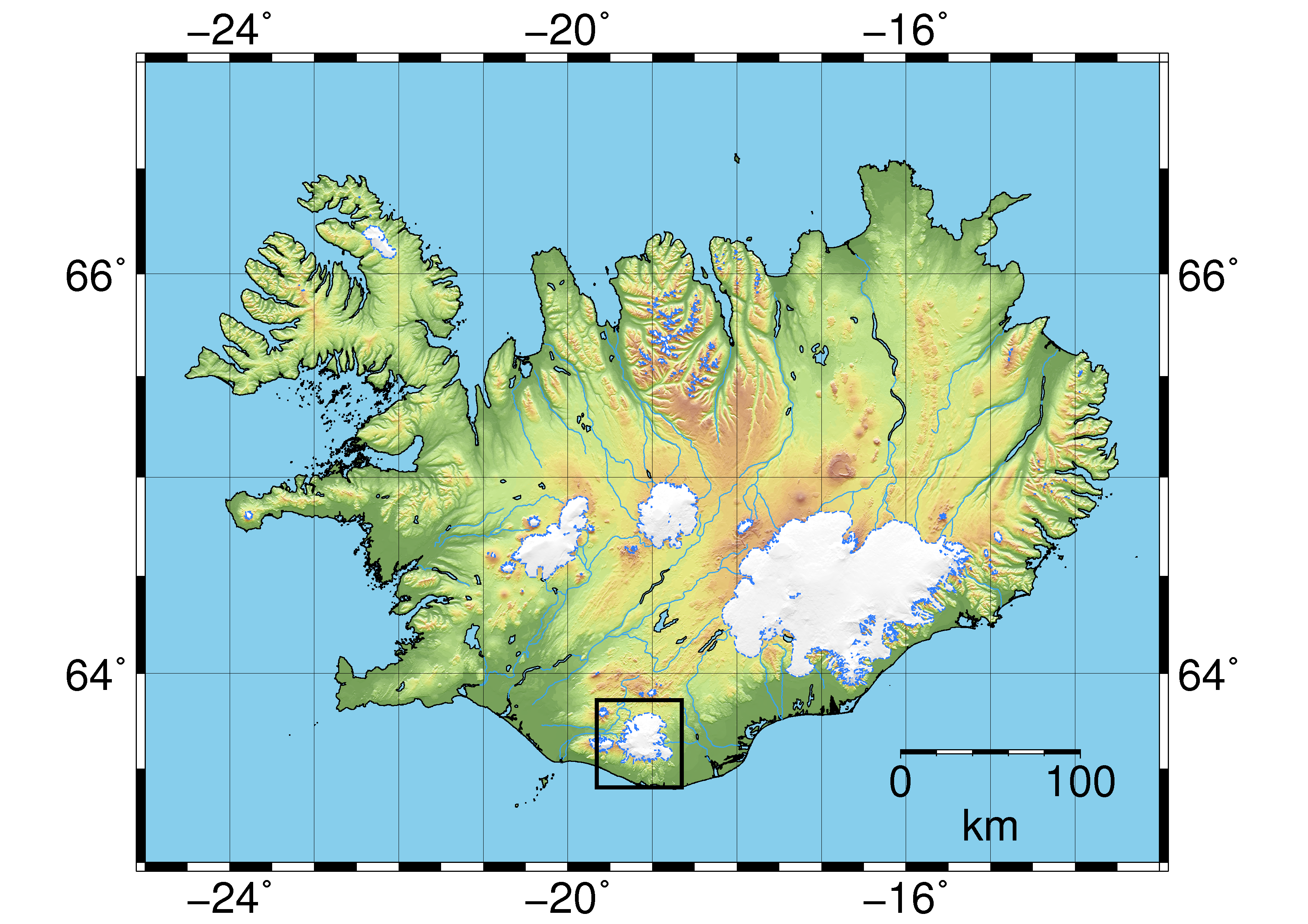}
\caption{A map of Iceland showing the location of Katla volcano (black solid box). The white areas mark glaciers.}
\label{fig:iceland}
\end{figure}

\end{document}